\documentclass[preprint,proceedings]{AMartinezSTDivBrmaa}

\SetYear{2016}
\SetConfTitle{XV Latin American Regional IAU Meeting LARIM}

\title{LAGO distributed network of data repositories}

 \author{H. Asorey\altaffilmark{1}, A. Mart\'inez-M\'endez\altaffilmark{2}, L.A. N\'u\~nez\altaffilmark{3,4} and A. Valbuena-Delgado\altaffilmark{3} for the LAGO Collaboration\altaffilmark{5}
  }

  \altaffiltext{1}{Laboratorio Detecci\'on de Part\'iculas y Radiaci\'on, Centro At\'omico Bariloche \& Instituto Balseiro, Bariloche, Argentina}

  \altaffiltext{2}{Escuela de Ingenier\'ia de Sistemas, Universidad Industrial de Santander, Bucaramanga, Colombia.}
  
  \altaffiltext{3}{Escuela de F\'isica, Universidad Industrial de Santander, Bucaramanga, Colombia.}

  \altaffiltext{4}{Departamento de F\'isica, Universidad de Los Andes, M\'erida, Venezuela.}
  
  \altaffiltext{5}{The Latin American Giant Observatory (LAGO), http://lagoproject.org, see the full list of members and institutions at http://lagoproject.org/collab.html}

\suppressfulladdresses

\listofauthors{H. Asorey}
\listofauthors{A. Mart\'inez-M\'endez}
\listofauthors{L. A. N\'{u}\~{n}ez}
\listofauthors{A. Valbuena-Delgado}
\listofauthors{Asorey, H.}
\listofauthors{Mart\'inez-M\'endez, A.}
\listofauthors{N\'{u}\~{n}ez, L. A.}
\listofauthors{Valbuena-Delgado, A. }
\addkeyword{Data Repository}
\addkeyword{Open Data}
\addkeyword{Data preservation}

\begin{document}
\maketitle 

\boldabstract{We describe a set of tools, services and strategies of the Latin American Giant Observatory (LAGO) data repository network, to implement  Data Accessibility, Reproducibility and Trustworthiness.}

The Latin American Giant Observatory, (LAGO), is an extended continental astroparticle observatory oriented to basic research on: the Extreme Universe, Space Weather, and Atmospheric Radiation, with singles and small arrays particle detectors, covering a huge range  of geomagnetic rigidity cutoffs and atmospheric absorption/reaction levels\,\citep{AsoreyEtal2016a}. 

Unlike other instruments where data flows from only one place to a network of data repositories, in LAGO each site preserves, catalogs and generates data locally which, referrers not only to raw data but also to data produced during the analysis and/or simulation of cosmic rays phenomena.   Dspace provides basic functionality for storing and retrieving of digital content with a straightforward adaptability for non-native types of contents and metadata schemes, supporting two interoperability protocols: OAI-PMH (Open Archive Initiatives Protocol for Metadata Harvesting) and SWORD (Simple Webservice Offering Repository Deposit)\citep{SmithEtal2003,LewisDeCastroJones2012}. We have overcome one of the most important DSpace limitations: its inability to upload/download multiple records, developing a script to ingest data profiting from the some DSpace capabilities\citep{AsoreyEtAl2015A}. This data repository network will also be useful for the solar physics and space climatology communities.

\begin{figure}[!t]
\centering
  \includegraphics[width=4cm]{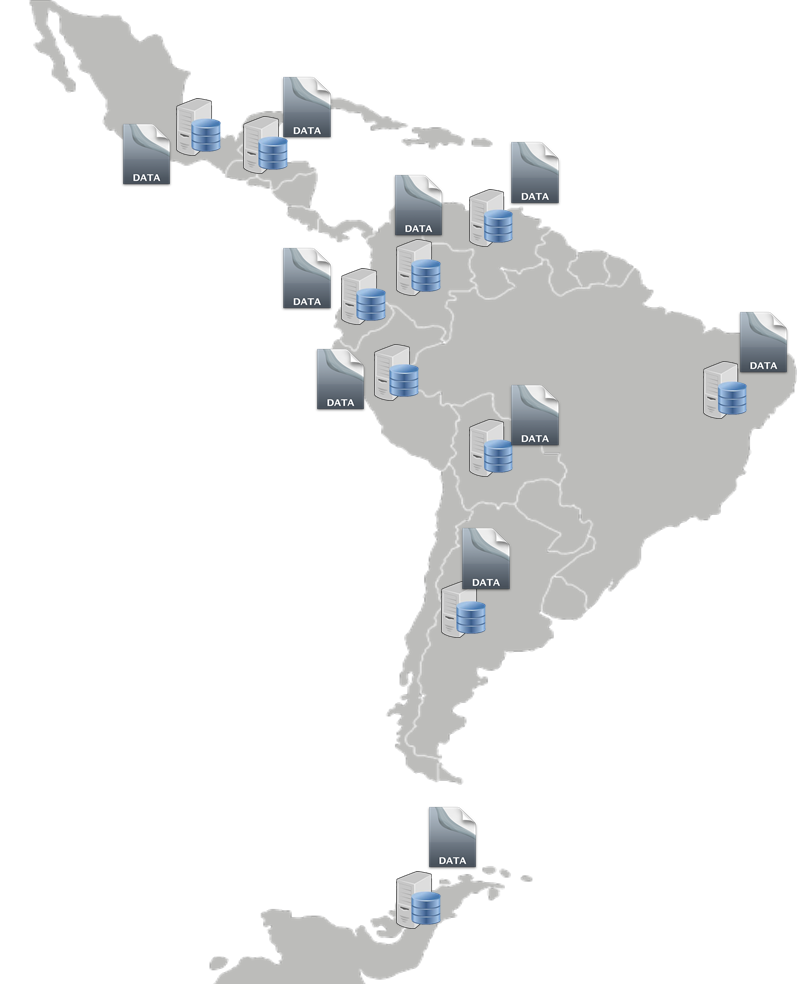}
  \caption{Continental LAGO data repository network.}
  \label{fig:simple}
\end{figure}

The Data Accessibility, Reproducibility and Trustworthiness (DART) initiative was launched by CHAIN-REDS (Coordination and Harmonisation of Advanced e-infrastructure for Research and Education Data Sharing), an European Commission co-funded project focused on promoting and supporting technological and scientific collaboration across different communities in various continents. 
GRNET PID (European Persistent Identifiers Consortium) service enables the allocation, management and resolution of PIDs and has been employed to ensure the data persistence and reproducibility of the experiments.  Part identifiers can compute an unlimited number of handles on the fly, without requiring registering each separately. \citep{BarberaEtal2014B}.

This project has been partially funded by VIE Universidad Industrial de Santander.

\bibliographystyle{unsrtnat}
\bibliography{/Users/luisnunez/Dropbox/MisDocumentos/Publicaciones/BibTex/BiblioLN170101}

\begin{thebibliography}
\expandafter\ifx\csname natexlab\endcsname\relax\def\natexlab#1{#1}\fi
\expandafter\ifx\csname href\endcsname\relax
  \def\href#1#2{}\fi
\expandafter\ifx\csname urllinklabel\endcsname\relax
  \def\urllinklabel{[LINK]}\fi
\expandafter\ifx\csname adsurllinklabel\endcsname\relax
  \def\adsurllinklabel{[ADS]}\fi

\bibitem[{Asorey {et~al.}(2015)Asorey, Cazar-Ram\'irez, Mayo-Garc\'ia,
  N\'u\~nez, Rodr\'iguez-Pascual, Torres-Ni\~no, \& the
  LAGO~Collaboration}]{AsoreyEtAl2015A}
Asorey, H., Cazar-Ram\'irez, D., Mayo-Garc\'ia, R., N\'u\~nez, L.,
  Rodr\'iguez-Pascual, M., Torres-Ni\~no, L., \& the LAGO~Collaboration. 2015,
  in The 34th International Cosmic Ray Conference, Vol. PoS(ICRC2015), 672


\bibitem[{Asorey {et~al.}(2016)Asorey, N{\'u}{\~n}ez, Su{\'a}rez-Dur{\'a}n,
  Torres-Ni{\~n}o, Rodr{\'\i}guez-Pascual, Rubio-Montero, \&
  Mayo-Garc{\'\i}a}]{AsoreyEtal2016a}
Asorey, H., N{\'u}{\~n}ez, L., Su{\'a}rez-Dur{\'a}n, M., Torres-Ni{\~n}o, L.,
  Rodr{\'\i}guez-Pascual, M., Rubio-Montero, A., \& Mayo-Garc{\'\i}a, R. 2016,
  in Cluster, Cloud and Grid Computing (CCGrid), 2016 16th IEEE/ACM
  International Symposium on, IEEE, 707--711


\bibitem[{Barbera {et~al.}(2014)Barbera, Becker, Carrubba, Inserra,
  Jalife-Villal\'on, Kanellopoulos, Koumantaros, Mayo-Garc\'ia, N\'u\~nez,
  Prnjate, Ricceri, Rodr\'iguez-Pascual, Rubio-Montero, Ruggieri, \&
  Project}]{BarberaEtal2014B}
Barbera, R., Becker, B., Carrubba, C., Inserra, G., Jalife-Villal\'on, S.,
  Kanellopoulos, C., Koumantaros, K., Mayo-Garc\'ia, R., N\'u\~nez, L.,
  Prnjate, O., Ricceri, R., Rodr\'iguez-Pascual, M., Rubio-Montero, A.,
  Ruggieri, F., \& Project, C.-R. 2014, in ANAIS DAS SESS{\~O}ES TEM{\'A}TICAS
  E P{\^O}STERS, 166


\bibitem[{Lewis {et~al.}(2012)Lewis, de~Castro, \&
  Jones}]{LewisDeCastroJones2012}
Lewis, S., de~Castro, P., \& Jones, R. 2012, D-Lib Magazine, 18


\bibitem[{Smith {et~al.}(2003)Smith, Barton, Bass, Branschofsky, McClellan,
  Stuve, Tansley, \& Walker}]{SmithEtal2003}
Smith, M., Barton, M., Bass, M., Branschofsky, M., McClellan, G., Stuve, D.,
  Tansley, R., \& Walker, J.~H. 2003, D-lib magazine, 9


\end{thebibliography}

\end{document}